\documentclass[aps,prb,twocolumn,superscriptaddress,nofootinbib, showemail]{revtex4-2}
\usepackage[utf8]{inputenc}
\usepackage[T1]{fontenc}
\usepackage{amsmath,amssymb,amsfonts}
\usepackage{graphicx}
\usepackage{bm}
\usepackage{xcolor}
\usepackage[colorlinks=true,citecolor=blue!70!black,linkcolor=blue!70!black,urlcolor=blue!70!black,breaklinks=true, hypertexnames=false]{hyperref}

\makeatletter
\newenvironment{widetextwithoutlines}{%
  \par\ignorespaces
  \setbox\widetext@top\vbox{}
  \setbox\widetext@bot\hb@xt@\hsize{}
  \onecolumngrid
  \dimen@\ht\widetext@top\advance\dimen@\dp\widetext@top
  \cleaders\box\widetext@top\vskip\dimen@
  \let\set@footnotewidth\set@footnotewidth@ii
}{%
  \par
  \setbox\widetext@bot\vbox{}
  \dimen@\ht\widetext@bot\advance\dimen@\dp\widetext@bot
  \cleaders\box\widetext@bot\vskip\dimen@
  \twocolumngrid\global\@ignoretrue\@endpetrue
}
\makeatother

\begin{document}

\title{Effect of the local field and dipole–dipole interaction on the spontaneous ordering of dipole moments in bismuth monolayers with an orthorhombic structure}

\author{D. V. Ponkratova}
\affiliation{Moscow Institute of Physics and Technology, Dolgoprudny, 141701 Russia}
\affiliation{Kotelnikov Institute of Radioengineering and Electronics of Russian Academy of Sciences, Moscow, 125009 Russia}
\affiliation{HSE University, Moscow, 101000 Russia}

\author{I. V. Zagorodnev}
\affiliation{Moscow Institute of Physics and Technology, Dolgoprudny, 141701 Russia}
\affiliation{Kotelnikov Institute of Radioengineering and Electronics of Russian Academy of Sciences, Moscow, 125009 Russia}
\affiliation{HSE University, Moscow, 101000 Russia}

\author{V. V. Enaldiev}\email{enaldiev.vv@mipt.ru}
\affiliation{Moscow Institute of Physics and Technology, Dolgoprudny, 141701 Russia}
\affiliation{Kotelnikov Institute of Radioengineering and Electronics of Russian Academy of Sciences, Moscow, 125009 Russia}

\begin{abstract}
Using two complementary approaches, the instability of bismuth monolayers with an orthorhombic structure with respect to the emergence of spontaneous electric dipole moments at the crystal lattice sites has been studied. The first approach, based on determining the dipole moments of lattice sites through the polarizability of bismuth atoms and taking into account the local Lorentz–Lorenz field, suggests the possibility of the existence of several antiferroelectric and ferroelectric phases in orthorhombic bismuth monolayers. Using the second approach, the vibrational spectrum of a nonpolar symmetric lattice has been analyzed, and two types of soft optical modes corresponding to the ferroelectric and antiferroelectric instabilities of the monolayers have been revealed. The relationship of the Born effective charges to the polarization direction in the ferroelectric phase, as well as their role in the reduction of the frequency of the polar optical phonon, which characterizes the lattice deformation in the ferroelectric phase due to the dipole–dipole interaction, has been shown.
\end{abstract}

\maketitle

\section*{INTRODUCTION}
The emergence of spontaneous electric polarization without an external electric field, called ferroelectricity, was first demonstrated by J. Valasek \cite{Valasek1921} and until recently was associated only with chemically complex compounds such as PbZr$_x$Ti$_{1-x}$O$_3$ \cite{Jaffe1954}, BaTiO$_3$ \cite{Acosta2017} and molecular crystals \cite{Wang2024}, where the electric polarization arises when cations are displaced relative to anions.

\begin{figure}[t]
\centering
\includegraphics[width=1.\columnwidth]{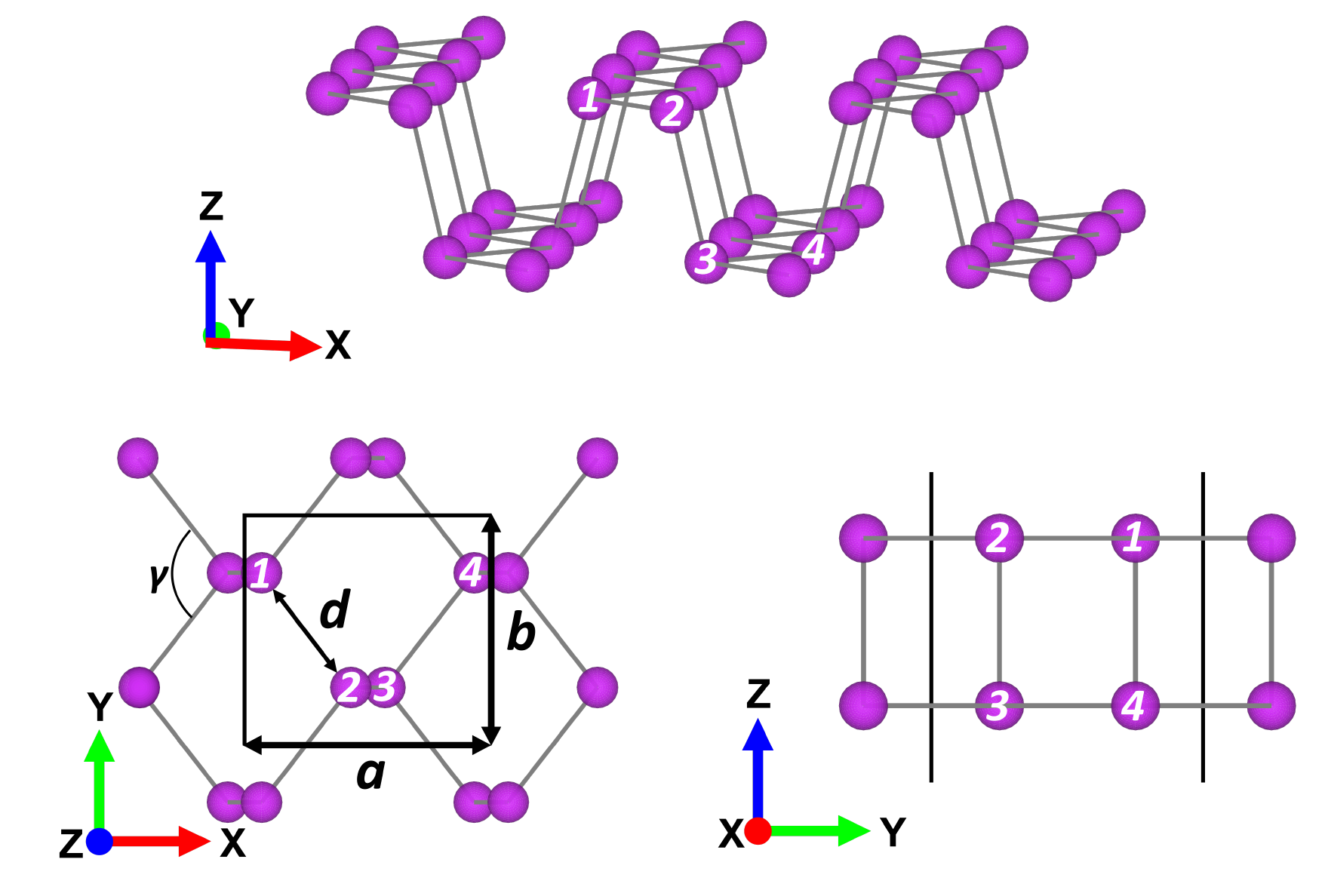}
\caption{Lattice structure of an orthorhombic bismuth monolayer in the nonpolar (paraelectric) phase. The numbers indicate the sublattice sites belonging to one unit cell.}
\label{fig1}
\end{figure}

However, a ferroelectric transition in atomically thin monoatomic crystals consisting only of bismuth atoms ordered in an orthorhombic lattice (see Fig.~\ref{fig1}), where the concept of cation/anion loses its meaning due to the identical valence of each atom, has been recently discovered in \cite{Gou2023}. Moreover, the electric polarization vector in the ferroelectric phase of such two-dimensional crystals lies in the plane of the layer, which is qualitatively consistent with the concept of amplification of the local electric field of a two-dimensional lattice of dipoles uniformly polarized in the plane \cite{Mikhailov2013}.

Although it is known from  ab initio calculations that the ferroelectric phase in monoatomic monolayers with an orthorhombic structure consisting of Group V elements \cite{Xiao2018} or Group IV monochalcogenides (such as lead sulfide) \cite{Sutter2021} is associated with a lattice deformation dictated by the softness of the corresponding out-of-plane optical mode of lattice vibrations (ZO) \cite{Akturk2016} (see Fig.~\ref{fig3}), the physical reason for its softening remains unclear. At the same time, it is well known that the frequency of polar phonons corresponding to lattice deformation in the ferroelectric phase can decrease due to the competition between short-range interatomic forces and long-range dipole-dipole forces \cite{Murzin1968}.

In this work, we analyze the types of ordering of electric dipoles at the sites of the orthorhombic lattice of bismuth monolayers, arising due to the local Lorentz–Lorenz field, for which the parameters are the microscopic atomic polarizability tensors and the lattice constants themselves. It is shown that several variants of ferroelectric and antiferroelectric orders, whose boundaries with the unpolarized (paraelectric) phase are located in an experimentally realistic range of parameters, exist for both the isotropic polarizability of bismuth atoms and the  ab initio calculated polarizability (see Fig.~\ref{fig2}). It is also shown that the ZO soft optical mode in the phonon spectrum, which stimulates the ferroelectric transition, wins the competition with the chiral (C) soft optical mode, which promotes the formation of the antiferroelectric order because of a significant decrease in its frequency due to the dipole–dipole interaction induced by the dipole moment of the ZO mode.

\section{SPONTANEOUS ORDERING OF DIPOLE MOMENTS}

In the nonpolar phase, an orthorhombic bismuth monolayer is characterized by a rectangular Bravais lattice $\mathbf{R}_n = n_1\mathbf{a} + n_2\mathbf{b}$, which is specified by the basis vectors $\mathbf{a} = (a, 0)$ (along the ``armchair'' crystallographic direction parallel to the $x$ axis) and $\mathbf{b} = (0, b)$ (along the ``zigzag'' crystallographic direction parallel to the $y$ axis), and by the position vectors $\bm{\tau}_s$ ($s = 1,2,3,4$) of four atoms in the unit cell, spaced from each other by the same distance $d$ (see Fig.~\ref{fig1}). In this work, we specified these vectors as
$\bm{\tau}_1=d(0,\sin(\gamma/2),\cos\beta)$, $\bm{\tau}_2=d(\cos(\gamma/2),0,\cos\beta)$, $\bm{\tau}_3=d(\cos(\gamma/2)+(1/2)\sin\beta,0,0)$, $\bm{\tau}_4=d(2\cos(\gamma/2)+(1/2)\sin\beta,\sin(\gamma/2),0)$, 
where $\gamma = 2\arcsin[b/(2d)]$ is the angle between the covalent bonds of the nearest atoms and $\beta = \arcsin[a/d - 2\cos(\gamma/2)]$. Considering the monolayer as an array of atoms with a given polarizability, we find the dipole moment $\mathbf{p}^{(s)}$ of each atom, which is induced by the sum of the external uniform field $\mathbf{E}^{\text{ext}}$ and the field $\mathbf{E}^{\text{ind}}$ created by the induced dipole moments of the remaining atoms of the lattice.

The field of a periodic array of dipoles for a bismuth atom at the site $\mathbf{R}_n + \bm{\tau}_s$ can be expressed in the form
\begin{equation}
E_{i}^{\mathrm{ind}}(\mathbf{R}_n + \bm{\tau}_s) = \sum_{s' = 1,2,3,4}\Pi_{ij}^{ss'}(\mathbf{q}\to 0)p_j^{(s')},
\label{eq1}
\end{equation}
where summation over repeating subscripts $i,j = x,y,z$ is implied and
\begin{equation}
\Pi_{ij}^{ss'}(\mathbf{q}) = \frac{1}{\varepsilon}\left[\frac{\partial^2}{\partial x_i\partial x_j}\sum_n\frac{e^{-i\mathbf{q}(\mathbf{R}_n - \mathbf{R}_{n'} + \bm{\tau}_{ss'})}}{\left|\mathbf{r} - \mathbf{R}_{n} + \mathbf{R}_{n'} - \bm{\tau}_{ss'}\right|_{\mathbf{r}\to 0}}\right].
\label{eq2}
\end{equation}
Here and below, the prime at the summation sign means that the term with $n=n'$ for the coinciding subblattice indiices $s=s'$, i.e., the self-interaction term, is excluded. In Eq.~\eqref{eq2}, where $\varepsilon$ is the static permittivity of the monolayer environment and  $\bm{\tau}_{ss'} = \bm{\tau}_s - \bm{\tau}_{s'}$, it is taken into account that the field of a unit dipole $\bf p$
at the point $\bm{r}=0$ can be rewritten as $\nabla_{\bm{r}}({\bf p}\cdot\nabla_{\bm{r}} |\bm{r}|^{-1})$ (where $\nabla_{\bm{r}}=\partial/\partial\bm{r}$). To calculate the sum in Eq.~\eqref{eq2}, we use the Ewald summation method \cite{Ewald1921} (see Section S2 in the supplementary material).

Thus, the dipole moment of the atom at the site $\mathbf{R}_n + \bm{\tau}_s$ is given by
\begin{equation}
p_i^{(s)} = \alpha_{ij}^{(s)}\left[E_j^{\mathrm{ext}} + E_j^{\mathrm{ind}}(\mathbf{R}_n + \bm{\tau}_s)\right],
\label{eq3}
\end{equation}
which, using Eq.~\eqref{eq1}, can be rewritten as
\begin{equation}
\sum_{s' = 1,2,3,4}\left[\alpha_{ij}^{(s)-1}\delta_{ss'} - \Pi_{ij}^{ss'}(\mathbf{q}\to 0)\right]p_j^{(s')} = E_i^{\mathrm{ext}},
\label{eq4}
\end{equation}
where $\alpha_{ij}^{(s)}$ is the polarizability tensor of the bismuth atom in the $s$th sublattice and $\delta_{ss'}$ is the Kronecker delta. For a given external field, system (4) determines the dipole moments and, consequently, the polarization of the monolayer $\mathbf{P} = \Omega^{-1}\sum_{s=1,2,3,4}^4 \mathbf{p}^{(s)}$ as a continuous medium ($\Omega$ is the volume of the unit cell of the monolayer). However, even in zero external field $E_i^{\mathrm{ext}} = 0$, the homogeneous system (4) has nontrivial solutions for $\mathbf{p}^{(s)}$ at the values of the parameters that vanish its determinant. The arising solutions can be separated into two types characterized by the ferroelectric ($\mathbf{P}\neq 0$) and antiferroelectric ($\mathbf{P} = 0$ for $\mathbf{p}^{(s)}\neq 0$) orders. The former type is accompanied by a divergence of the monolayer susceptibility (``polarization catastrophe''), while the latter type corresponds to a finite positive susceptibility, as in the nonpolar phase \cite{Kittel1951}.
\begin{figure}[!t]
\centering
\includegraphics[width=1.0\columnwidth]{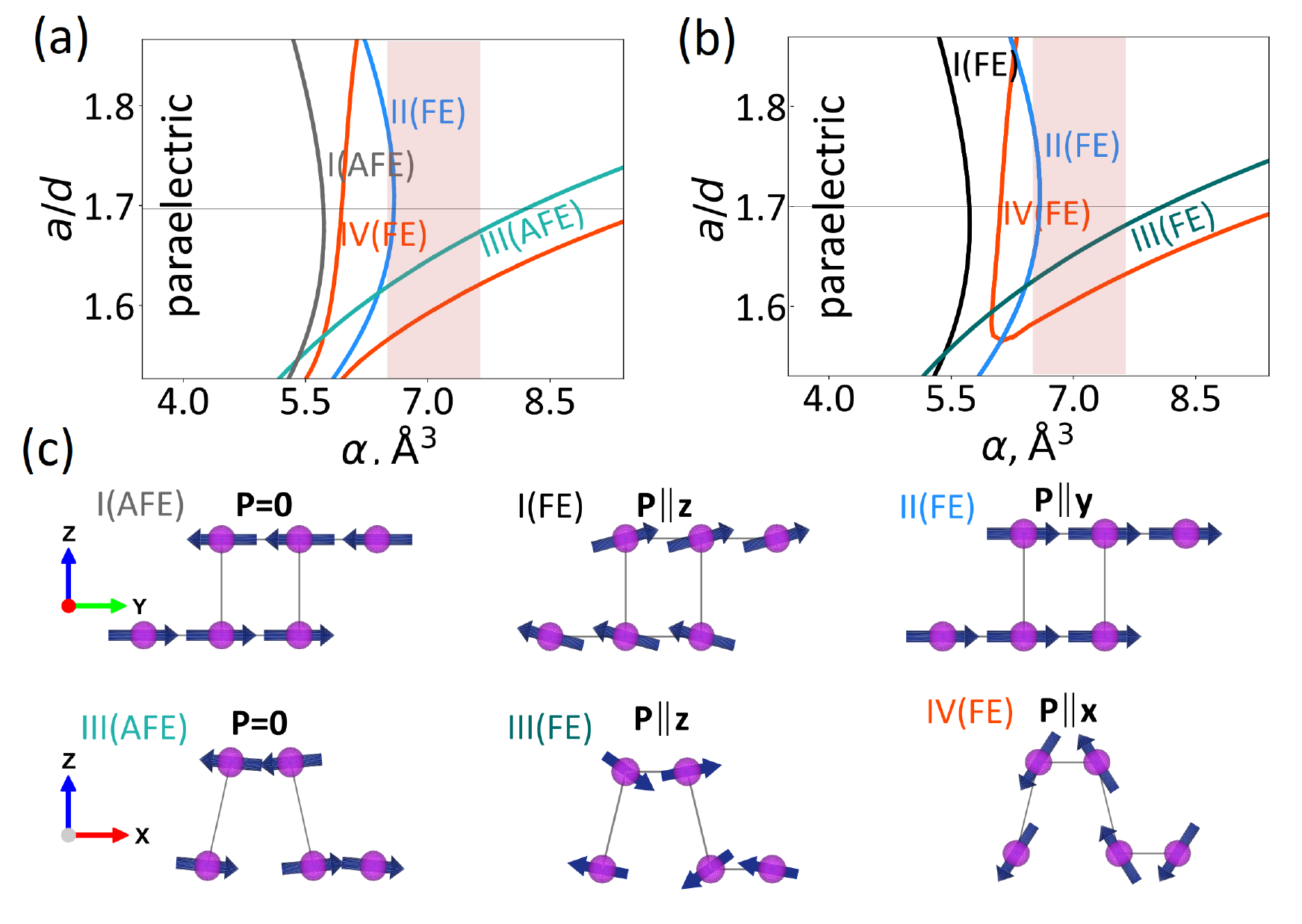}
\caption{(a, b) Parametric dependences of $\alpha$ on $a/d$ for the boundaries between the nonpolar (paraelectric) and spontaneously polarized phases with the distributions of dipole moments calculated with the parameters $d = 2.8\ \text{\AA}$, $b = 4.42\ \text{\AA}$, and (a) $\alpha_{ij}^{1,2,3,4} = \alpha \delta_{ij}$ and (b) $\alpha_{ij}^{1,2,3,4}$ given by Eqs.~(5)--(8). The black horizontal line corresponds to $a = 4.75\ \text{\AA}$ from the experiment reported in \cite{Gou2023,Gou2020}. The range of relevant $\alpha$ values is shaded in pink. (c) Distributions of dipole moments corresponding to the phase boundaries in (a, b).}
\label{fig2}
\end{figure}
Figure~\ref{fig2} shows the lines in the $(\alpha, a/d)$ plane, where $\alpha$ is the diagonal component of the polarizability tensor of bismuth atoms, corresponding to the zeros of the determinant of system (4), along which nontrivial solutions arise.

Figure~\ref{fig2}(a) corresponds to the isotropic polarizability $\alpha$ for all sublattices, as for a free atom. At small polarizabilities $\alpha\lesssim5\,$\AA$^3$, the sublattices have zero dipole moment, which corresponds to the paraelectric phase. By increasing the polarizability at a fixed parameter $a/d$, it is possible to achieve the intersection of the boundaries of one of the antiferroelectric or ferroelectric orders with the distribution of the spontaneous dipole moment of the sublattices, as shown in the lower insets of Fig.~\ref{fig2}. The experimental lattice parameters correspond to the horizontal black line \cite{Gou2020}, according to which, as $\alpha$ increases, the monolayer first passes into the antiferroelectric phase I(AFE), with the opposite directions of the local dipole moments in the two sublayers ($p_y^{(1)}=p_y^{(2)}=-p_y^{(3)}=-p_y^{(4)}\neq0$). At large $\alpha$ values, a transition either to another antiferroelectric phase III(AFE) or to one of the ferroelectric phases IV(FE) $(\mathbf{P}\| x)$ and II(FE) $(\mathbf{P}\| y)$ is possible. It should be noted that the boundaries of all phases for the considered lattice parameters are located in the physically relevant range of atomic polarizabilities $6.5\,\text{\AA}^{3}\lesssim\alpha\lesssim 7.7 \,\text{\AA}^{3}$ (highlighted in pink in Figs.~\ref{fig2}(a), \ref{fig2}(b)) \cite{Maroulis2006,Schwerdtfeger2019}, which indicates the possibility of competition between them.

An estimate of the polarizability tensors of bismuth atoms in an orthorhombic lattice from ab initio calculations reveals their difference from the isotropic case discussed above (see Section S1 in the supplementary material) due to the hybridization of the valence electron orbitals responsible for the formation of chemical bonds and simultaneously determining the polarization response of the monolayer. For comparison with the isotropic case, we parameterize $\alpha_{ij}^{(s)}$ as follows:
\begin{align} 
\alpha_{xx}^{(s)}&=\alpha_{yy}^{(s)}=\alpha_{zz}^{(s)}=\alpha,\quad s=1,2,3,4; \label{eq5}\\
\alpha_{xz}^{(1)}&=-\alpha_{xz}^{(2)}=-\alpha_{xz}^{(3)}=\alpha_{xz}^{(4)}=-0.4\,\text{\AA}^3, \label{eq6}\\
\alpha_{zx}^{(1)}&=\alpha_{zx}^{(2)}=-\alpha_{zx}^{(3)}=-\alpha_{zx}^{(4)}=-1.4\,\text{\AA}^3, \label{eq7}\\
\alpha_{yz}^{(1)}&=\alpha_{yz}^{(2)}=-\alpha_{yz}^{(3)}=-\alpha_{yz}^{(4)}=4.8\,\text{\AA}^3, \label{eq8}
\end{align}
where all diagonal elements of the tensor are equated for simplicity. Figure~\ref{fig2}(b) shows the phase boundaries calculated from the conditions of vanishing the determinant of system \eqref{eq4} taking into account Eqs.~\eqref{eq5}--\eqref{eq8}.

The inclusion of the off-diagonal components in $\alpha_{ij}^{(s)}$ does not lead to a significant change in the positions of the phase boundaries, but boundaries I(FE) and III(FE) now correspond to the uncompensated dipole moment of the cell with spontaneous polarization ${\bf P}||z$. Nevertheless, they should be distinguished, since the monolayer susceptibility at the boundary I(FE) remains finite and positive, while the susceptibility at the boundary III(FE) diverges, as for the ferroelectric phases II(FE) and IV(FE).

Thus, the inclusion of the local field indicates the possibility of realizing several variants of ferroelectric and antiferroelectric orders whose boundaries, being close to each other, can interchange in the parametric space depending on the type of the atomic polarizability tensor. However, since the transition to any phase in a real orthorhombic monolayer must be accompanied by a structural transformation of the lattice, breaking the symmetry of the paraelectric phase, we next examine the phonon spectrum to determine soft optical phonon modes.

\section{SOFT PHONON MODES}

\begin{figure}[!t]
\centering
\includegraphics[width=1\columnwidth]{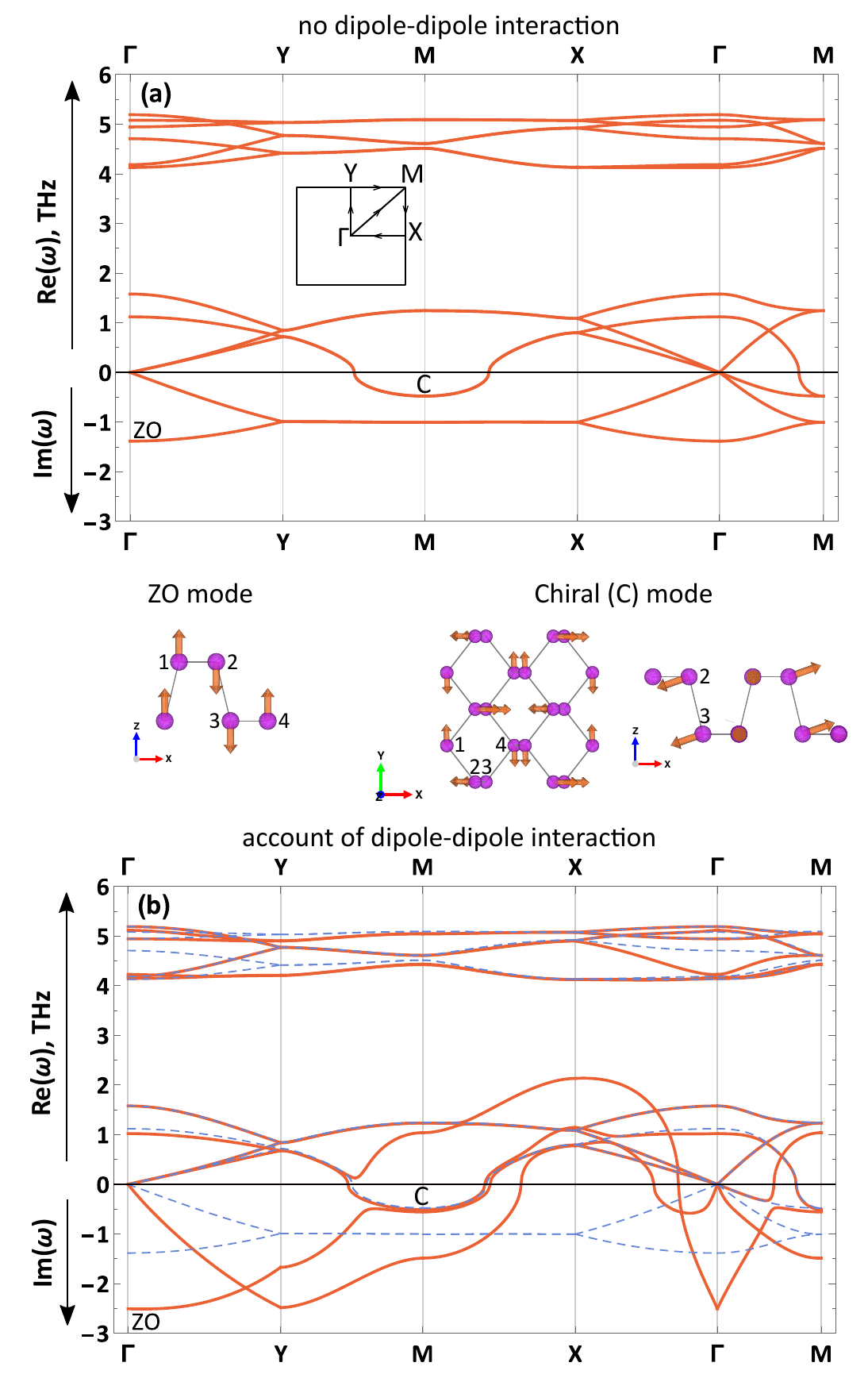}
\caption{(a) Phonon spectrum in the orthorhombic bismuth monolayer obtained from Eq.~(11) in the nearest-neighbour approximation disregarding the Born charges $(Z_{ji}^{s} = 0)$ along the path in the Brillouin zone shown in the inset. (b) Same spectrum taking into account the Born charges $|Z_{xx}^{s}| = 1.5e$ at $\epsilon = 1$ (the blue dashed line corresponds to the spectrum in panel (a)). The arrows in the inset indicate the directions of sublattice displacements for the ZO mode at $\mathbf{q}_{\mathbf{M}} = 0$ and the C mode at $\mathbf{q}_{\mathbf{M}} = (\pi/a, \pi/b)$.}
\label{fig3}
\end{figure}

To study the phonon spectrum in an orthorhombic bismuth monolayer, we use the Lagrangian
\begin{multline}\label{eq9}
		L =\sum_{n,s} \frac{M\dot{u}^{s}_{i}({\bf R}_n)\dot{u}^{s}_{i}({\bf R}_{n})}{2}-\\
		-\frac{1}{2}\sum_{n,n',s,s'}\Phi_{ij}^{ss'}({\bf R}_n-{\bf R}_{n'})u^{s}_{i}({\bf R}_n)u^{s'}_{j}({\bf R}_{n'})+\\ +\sum_{n,s}Z^{(s)}_{ji}u^{s}_{i}({\bf R}_n)E_j({\bf R}_n+\bm{\tau}_s),
\end{multline}
where $u_{i}^s({\bf R}_n)$ and $\dot{u}_{i}^{s}({\bf R}_n)$ are the displacement and velocity of an atom in the $i$ direction at the lattice site $\mathbf{R}_n + \bm{\tau}_s$, respectively, $M$ is the mass of the bismuth atom, $\Phi_{ij}^{ss'}(\mathbf{R}_n - \mathbf{R}_{n'})$ is the force constant matrix, $Z_{ji}^{s}$ is the Born charge tensor for the $s$th sublattice, and $E_j(\mathbf{R}_n + \bm{\tau}_s)$ is the electric field created at the site $\mathbf{R}_n + \bm{\tau}_s$ when all other atoms are displaced from their equilibrium positions. The first two terms in Eq.~\eqref{eq9} are necessary to describe the harmonic vibrations of the monolayer atoms, while the last term describes the occurrence of the electric polarization during optical vibrations. By analogy with Eq.~\eqref{eq1}, the electric field caused by such polar phonons is given by
\begin{equation}
E_j(\mathbf{R}_n + \bm{\tau}_s) = \sum_{n',s'}\Pi_{jk}^{ss'}(\mathbf{q}\to 0)Z_{kp}^{s'}u_p^{s'}(\mathbf{R}_{n'}),
\label{eq10}
\end{equation}
where $Z_{kp}^{s'}u_p^{s'}$ is the dipole moment of the site $\mathbf{R}_{n'} + \bm{\tau}_{s'}$. Despite the monoatomic composition of the crystal and the same number of bonds between nearest neighbors, the ab initio calculation indicates the presence of a single nonzero component of the Born charge tensor $Z_{xz}^{(1)}=-Z_{xz}^{(2)}=-Z_{xz}^{(3)}=Z_{xz}^{(4)}\approx1.5\,e$ (see Section S1 in the supplementary material), which corresponds to the appearance of crystal polarization along the crystallographic armchair axis (parallel to the $x$ axis) for the ZO optical phonon mode (see the inset of Fig.~\ref{fig3}).

Minimizing the Lagrangian (9) and passing to the Fourier components of the displacements, we obtain
\begin{eqnarray}\label{eq11}
		\omega^2u^{s}_{i}( {\bf q})=\qquad\qquad\qquad\qquad\qquad\qquad\qquad\qquad\qquad\\
		= \sum_{s'}\left[\widetilde{\Phi}^{ss'}_{ij}\left({\bf q}\right)u^{s'}_{j}( {\bf q}) - Z^{s}_{ji}Z^{s'}_{kp}\Pi^{ss'}_{jk}({\bf q})u^{s'}_{p}( {\bf q}) \right],\nonumber 
\end{eqnarray}
where $u_i^{(s)}( {\bf q})=N^{-1}\sum_{n}e^{-i[\omega t+{\bf q} ({\bf R}_n+\bm{\tau}_s)]}u^{s}_{i}({\bf R}_n)$, $\mathbf{q} = (q_x, q_y, 0)$ is the quasi-wave vector defined in the first Brillouin zone (see the inset of Fig.~\ref{fig3}(a)), $N$ is the number of unit cells in the crystal, and \mbox{$\widetilde{\Phi}^{ss'}_{ij}\left({\bf q}\right)=M^{-1}\sum_{n}e^{-i {\bf q}\left( {\bf R}_{n}+ {\bm{\tau}}_{s}- {\bm{\tau}}_{s'}\right)}\Phi_{ij}^{ss'}({\bf R}_n-\bm{\tau}_s+\bm{\tau}_{s'})$} is the dynamical matrix.

Next, to determine the structural instabilities of the lattice caused with dipole ordering, we analyze the phonon spectrum in the unpolarized paraelectric phase, which is experimentally realized at the high temperatures $T \geq 210$~K \cite{Gou2023}. To calculate the phonon spectrum, we used the nearest-neighbour approximation for the force constant matrix, which, taking into account mirror symmetry in the plane perpendicular to the zigzag direction (parallel to the $y$ axis) and inversion, is parameterized as follows (see Fig. S2 in the supplementary material):
    \begin{align}
        \widetilde{\Phi}_{ij}^{11/22}({\bf q}) = & \widetilde{\Phi}_{ij}^{44/33}({\bf q})=
        \begin{pmatrix}
        \Phi_{xx}^{11} & 0 & \pm\Phi_{xz}^{11} \\
        0 & \Phi_{yy}^{11}  &  0 \\
        \pm\Phi_{xz}^{11} & 0 & \Phi_{zz}^{11}
    \end{pmatrix}. \label{paramet11}\\ 
    \widetilde{\Phi}_{ij}^{12}({\bf q}) = & 
        \begin{pmatrix}
        \Phi_{xx}^{12} & \Phi_{xy}^{12} & 0 \\
        \Phi_{xy}^{12} & \Phi_{yy}^{12} & \Phi_{yz}^{12} \\
      0 &  -\Phi_{yz}^{12} & \Phi_{zz}^{12}
    \end{pmatrix}e^{-i{\bf q}\bm{\tau}_{12}}+\\
    +&\begin{pmatrix}
        \Phi_{xx}^{12} & -\Phi_{xy}^{12} & 0 \\
        -\Phi_{xy}^{12} & \Phi_{yy}^{12} & -\Phi_{yz}^{12} \\
      0 &  \Phi_{yz}^{12} & \Phi_{zz}^{12}
    \end{pmatrix}e^{-i{\bf q}(\bm{\tau}_{12}-{\bf b})},\label{paramet12}\nonumber \\
    \widetilde{\Phi}_{ij}^{13}({\bf q})  = &\, 0, \\
    \widetilde{\Phi}^{14}_{ij}({\bf q})=&
    \begin{pmatrix}
        \Phi_{xx}^{14} & 0 & -\Phi_{xz}^{11} \\
        0 & \Phi_{yy}^{14}  &  0 \\
        -\Phi_{xz}^{11} & 0 & \Phi_{zz}^{14}
    \end{pmatrix}e^{-i{\bf q}(\bm{\tau}_{14}+{\bf a})}, \label{paramet14}
    \end{align}
    \begin{align}
    \widetilde{\Phi}^{23}_{ij}({\bf q})=&\begin{pmatrix}
        \Phi_{xx}^{14} & 0 & \Phi_{xz}^{11} \\
        0 & \Phi_{yy}^{14}  &  0 \\
        \Phi_{xz}^{11} & 0 & \Phi_{zz}^{14}
    \end{pmatrix}e^{-i{\bf q}\bm{\tau}_{23}}, \label{paramet23} \\
    \widetilde{\Phi}_{ij}^{24}({\bf q}) = &\,0,
    \\
    \widetilde{\Phi}_{ij}^{34}({\bf q}) = & 
        \begin{pmatrix}
        \Phi_{xx}^{12} & -\Phi_{xy}^{12} & 0 \\
        -\Phi_{xy}^{12} & \Phi_{yy}^{12} & \Phi_{yz}^{12} \\
      0 &  -\Phi_{yz}^{12} & \Phi_{zz}^{12}
    \end{pmatrix}e^{-i{\bf q}\bm{\tau}_{34}}+\label{paramet34}\\
    +&\begin{pmatrix}
        \Phi_{xx}^{12} & \Phi_{xy}^{12} & 0 \\
        \Phi_{xy}^{12} & \Phi_{yy}^{12} & -\Phi_{yz}^{12} \\
      0 &  \Phi_{yz}^{12} & \Phi_{zz}^{12}
    \end{pmatrix}e^{-i{\bf q}(\bm{\tau}_{34}+{\bf b})}. \nonumber 
    \end{align}

The parameters that appear in Eqs.\eqref{paramet11}--\eqref{paramet34} and are obtained from ab initio calculations (see Section S3 in the supplementary material) for the experimental parameters of a symmetric nonpolar lattice are summarized in Table~\ref{tab1}. The remaining blocks of the dynamical matrix are obtained from the Hermiticity requirement.

\begin{table}[htbp]
\centering
\caption{Parameters of the dynamical matrix (in electronvolts per square angstrom) obtained from {\it ab initio} calculations.}
\label{tab1}
\begin{tabular}{|l|c|c|c|c|}
\hline
 & \multicolumn{1}{c|}{\(xx\)} & \multicolumn{1}{c|}{\(yy\)} & \multicolumn{1}{c|}{\(zz\)} & \multicolumn{1}{c|}{\(xz\)/\(xy\)/\(yz\)} \\
 \hline
\(\Phi^{11}\) & 8.348 & 11.505 & 9.1309 & 1.689 (\(xz\)) \\
\(\Phi^{12}\) & -3.74 & -5.22 & 0.409 & 4.54 (\(xy\)), -0.13 (\(yz\)) \\
\(\Phi^{14}\) & -0.868 &  -1.065 & -9.9489 & --- \\
\hline
\end{tabular}
\end{table}

Since the experimental lattice parameters do not correspond to the equilibrium position of the symmetric lattice in the ab initio calculations, branches with imaginary frequencies arise in the phonon spectrum (see Fig.~\ref{fig3}(a)), which is obtained by diagonalizing the system of Eqs.~\eqref{eq11} and \eqref{paramet11}--\eqref{paramet34} with the parameters from Table~\ref{tab1} even disregarding the dipole–dipole interaction ($\Pi_{ij}^{ss'}=0$). Since the symmetric lattice describing the unpolarized phase of the monolayer in the experiment reported in \cite{Gou2023} is stable above the ferroelectric transition (at temperatures $\geq 210$~K), the force constants should depend on the temperature. Although such a temperature dependence cannot be extracted from the ab initio calculations by varying only three of twelve parameters of the dynamical matrix, the stability of all branches of the spectrum can be achieved (see Section S4 in the supplementary material), without changing the dispersion of the remaining initially stable branches of the spectrum. Since the choice of the original or adjusted force constants does not change the main result of the soft mode analysis, we used the original values.

Two soft optical modes appear in the phonon spectrum (see Fig.~\ref{fig3}(a)). The first of them, ZO mode, corresponds to a sublattice shift perpendicular to the layer plane, $\bm{u}_{\rm ZO}^1({\bf q})=-\bm{u}_{\rm ZO}^2({\bf q})=-\bm{u}_{\rm ZO}^3({\bf q})=\bm{u}_{\rm ZO}^4({\bf q})=(0,0,0.5)$, with the minimum $\omega''$ value at $Z_{xz}^1=-Z_{xz}^2=-Z_{xz}^3=Z_{xz}^4$, it becomes clear that the corresponding instability is responsible for the emergence of the uniform ferroelectric order, characterized by the polarization $P_x = \sum_{s} Z_{xz}^{s} u_{z,ZO}^{s} \neq 0$ along the crystallographic armchair axis. Note that it is this direction of spontaneous polarization and lattice deformation from the nonpolar phase that was observed experimentally \cite{Gou2023}.

The second soft C mode, which arises at the corner of the Brillouin zone $\mathbf{q}_{\mathrm{M}} = (\pi/a, \pi/b)$, describes a combination of chiral and out-of-plane sublattice shifts (see Fig.~\ref{fig3}). Since the displacements of atoms in adjacent unit cells for a phonon with the wave vector $\mathbf{q}_{\mathrm{M}}$ occur in opposite directions along both axes and $u_{z,C}^{2}(\mathbf{R}_n) = u_{z,C}^{3}(\mathbf{R}_n) \neq 0$, this soft mode promotes the emergence of the antiferroelectric order with a unit cell consisting of four cells of the unpolarized monolayer.

Since only the component $Z_{xz}^{s}$ in the orthorhombic bismuth monolayer is not equal to zero, the inclusion of the dipole–dipole interaction in Eq.~\eqref{eq11} mainly modifies the spectrum of phonon modes with displacement polarization perpendicular to the plane of the layer, as shown in Fig.~\ref{fig3}(b). As expected \cite{Murzin1968}, the inclusion of the dipole interaction significantly reduces the frequencies of the ZO mode with $\mathbf{q}=0$ because dipoles oriented in the plane increase the local polarizing field \cite{Mikhailov2013}. Note that the inclusion of the dipole interaction for the force constants corresponding to the stable spectrum in the absence of the dipole interaction also leads to an imaginary frequency for the ZO mode (see Section S4 in the supplementary material).

On the contrary, the frequency of the chiral C mode does not change when the dipole–dipole interaction is taken into account, since this interaction induces opposite polarizations in neighbouring cells, which leads to compensation of the total electric field.

Thus, the polarity of oscillations in the ZO mode, arising due to the only non-zero component $Z_{xz}^{s}$ of the Born charge tensor, ensures a significant decrease in its imaginary frequency $\omega''$ compared to the C mode and, accordingly, makes the transition to the ferroelectric order more energetically favourable compared to the antiferroelectric one.

\section*{ACKNOWLEDGMENTS}
The study was conducted using the HSE University supercomputer complex. This work was supported by the Russian Science Foundation (project no. 24-72-10015).

\bibliographystyle{apsrev4-2}
\bibliography{main}

\clearpage

\begin{widetextwithoutlines}
\begin{center}
{\large\bfseries Supplementary Material to the article ``Effect of the local field and dipole–dipole interaction on the spontaneous ordering of dipole moments in bismuth monolayers with an orthorhombic structure''}
\end{center}
\vspace{1em}
\end{widetextwithoutlines}

\renewcommand{\thesection}{S\arabic{section}}
\setcounter{section}{0}
\renewcommand{\theequation}{S\arabic{equation}}
\setcounter{equation}{0}
\renewcommand{\thefigure}{S\arabic{figure}}
\setcounter{figure}{0}
\renewcommand{\thetable}{S\arabic{table}}
\setcounter{table}{0}

\section{Estimate of bismuth atom polarizabilities in orthorombic monolayers from first principles}
First-principles calculations were carried out within the density functional theory (DFT) formalism implemented in the Vienna Ab-initio Simulation Package (VASP)~[\hyperlink{ref:1}{1},\hyperlink{ref:2}{2}]. Local density (LDA) [\hyperlink{ref:3}{3}] and general gradient (GGA-PBE) [\hyperlink{ref:4}{4}] approximations were used
for exchange-correlation functional with a projector augmented-wave 
pseudopotentials [\hyperlink{ref:5}{5}], treating the Bi
6s 6p as valence electrons. The plane-wave
energy cutoff was 550 eV. Convergence achieved when
the total energy change between steps is less than 10$^{-7}$
eV. A dipole correction [\hyperlink{ref:6}{6}] was applied for account of homogeneous external
electric field. Spin-orbit coupling (SOC) effect was not included in all calculations. In all calculations we used experimental values of lattice parameters [\hyperlink{ref:7}{7}] $a=4.75$\,\AA, $b=4.42$\,\AA, and $d=2.8$\,\AA.

We determine atomic polarizability tensors, $\alpha_{ij}^{(s)}$, for atoms occupying each sublattice $s=1,2,3,4$ with the help of the following indentity:
\begin{equation} \label{eq:alpha}
    p_i^{(s)} = \alpha^{(s)}_{ij} E_j^{(s)},
\end{equation}
where $p_i^{(s)}$ is $i$-th ($i=x,y,z$) component of atomic dipole moment for sublattice  induced by the $j$-th component of local electric field $E_j^{(s)}$. Polarizability of each sublattices is calculated dividing monolayer unit cells into four Voronoi cells (VC) centered at the position $\bm{\tau}_s$ in such a way that sum of VC volumes equals unit cell volume and
\begin{equation}
    \int_{\rm VC_{s}}\left[\rho_{{\bf E^{\rm ext}}}(\bm{r})-\rho_{0}(\bm{r})\right]d^3r=0,\quad s=1,2,3,4,
\end{equation}
where $\rho_{{\bf E^{\rm ext}}}(\bm{r})$ and $\rho_{0}(\bm{r})$ are electron densities computed in DFT at finite and zero external electric field, respectively. Then, the dipole moments of atoms in a unit cell are evaluated as follows:
\begin{equation}\label{eq:p}
\bm{p}^{(s)} =\int_{\rm VC_{s}} \left[\rho_{{\bf E^{\rm ext}}}(\bm{r})-\rho_{0}(\bm{r})\right]\bm{r}d^3r.
\end{equation}

\begin{figure}
    \centering
    \includegraphics[width=1.0\linewidth]{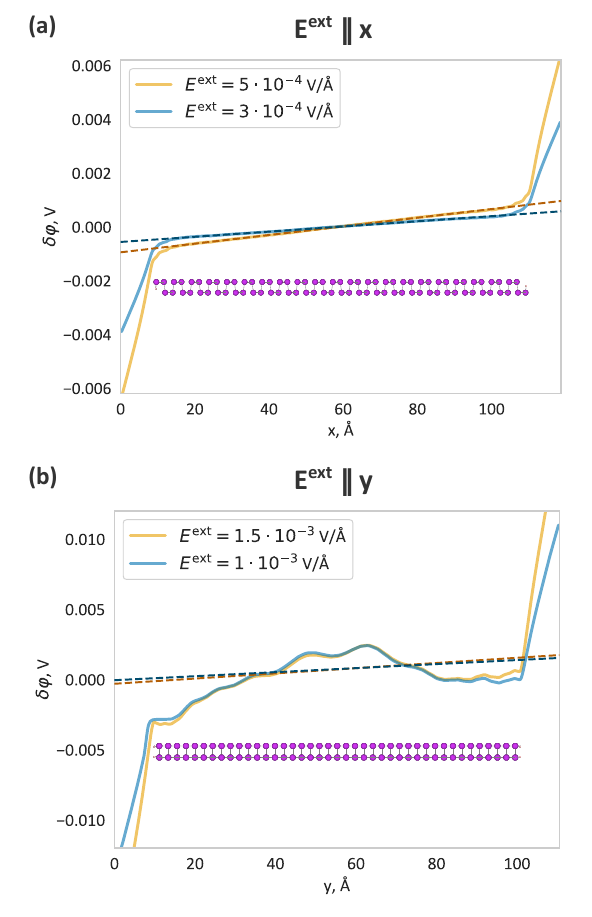}
    \caption{Planar averaged potential difference $\delta \varphi = \varphi_{{\bf E}^{\rm ext}}-\varphi_0$ for ${\bf E^{\rm ext}}||Ox$ (a) and ${\bf E^{\rm ext}}||Oy$ (b). The dashed lines indicate fits of slopes for the potential that were used to estimate local fields in VC for each sublattice. The insets schematically show the corresponding nanoribbon geometries. }
    \label{fig:potentials}
\end{figure}

Local electric field, $E_j^{(s)}$ were determined in different ways for in-plane and out-of-plane orientations of applied external field. For the in-plane direction of external field, we determined the local field via slope of total potential difference, $\varphi_{{\bf E}^{\rm ext}}-\varphi_0$ (shown in Fig. \ref{fig:potentials}), computed with finite, $\varphi_{{\bf E}^{\rm ext}}$, and zero, $\varphi_{0}$, external fields, for nanoribbons of width $88.4$~\AA{} (for $Ox||$ armchair axis) and $99.4$~\AA{} (for $Oy||$ zigzag axis) with in-plane vacuum gaps between periodic images being $22.1$~\AA{} and $19.3$~\AA{}, respectively, and out-of-plane gaps 20\,\AA. Note, that cutting bismuth ribbons from monolayer leads to undesirable metallic edge states in the gap. Therefore, to eliminate the edge states from the gap of bismuth ribbons we passivated them with hydrogen atoms that positions were relaxed  until convergence between steps of total energy were less than $10^{-6}$ eV and the forces acting on atoms were less than $10^{-3}$ eV/\AA. The relaxed distances of Bi–H bond were $1.84$\,\AA\, and $1.85$\,\AA\, for nanoribbons with armchair and zigzag edges, respectively.

For out-of-plane direction of ${\bf E^{\rm ext}}||Oz$ we calculated the local field as follows:
\begin{multline}\label{eq:E}
E_j^{(s)} =\\
\frac{-\int_{\text{VC}_s} \frac{\partial}{\partial x_j} \left[\varphi_{E^{\rm ext}_z}-\varphi_{0}-\int\frac{(\rho_{E^{\rm ext}_z}(\bm{r}')-\rho_{0}(\bm{r}'))}{|\bm{r} - \bm{r}'|} d^3\bm{r}'\right] \, \rho_0(\bm{r}) \, d^3\bm{r}}{\int_{\text{VC}_{s}} \rho_0(\bm{r}) \, d^3\bm{r}},
\end{multline}
where the last term in square brackets on the right-hand side excludes electric field produced by the electron density inside its VC (i. e. self-action).

For ${\bf E}^{\rm ext}$ along armchair ($||x$), zigzag ($||y$) and out-of-plane directions the Brillouin zone integration for corresponding primitive cells was performed on $k$‑point grids of $11 \times 1 \times 1$, $1 \times 11 \times 1$, and $11 \times 11 \times 1$, respectively.

As a result we obtained the following values for polarizability tensors (in \AA$^3$):
\begin{align}
\alpha^{(1)}_{ij} &= 
\begin{pmatrix}
    6.7 & -0.4 & -0.4 \\
    0.3 &  7.8 &  0.1 \\
   -1.4 &  4.8 &  6.4
\end{pmatrix}, \\
\alpha^{(2)}_{ij} &= 
\begin{pmatrix}
    6.7 &  0.4 &  0.4 \\
   -0.3 &  7.8 &  0.1 \\
   -1.4 &  4.8 &  6.4
\end{pmatrix}, \\
\alpha^{(3)}_{ij} &= 
\begin{pmatrix}
    6.7 & -0.4 &  0.4 \\
    0.3 &  7.8 & -0.1 \\
    1.4 & -4.8 &  6.4
\end{pmatrix}, \\
\alpha^{(4)}_{ij} &= 
\begin{pmatrix}
    6.7 &  0.4 & -0.4 \\
   -0.3 &  7.8 & -0.1 \\
    1.4 & -4.8 &  6.4
\end{pmatrix}.
\end{align}

\section*{Calculation of Born effective charges}
The Born effective charge tensors $Z^{(s)}_{ij}$ were calculated within DFT. The exchange–correlation interaction was described by the GGA–PBE functional. The plane‑wave energy cutoff is 600 eV. The Brillouin zone of the primitive monolayer cell was sampled with a $24 \times 24 \times 1$ $k$‑point mesh. The energy convergence criterion was $10^{-9}$ eV.

The obtained Born charge tensors for the four sublattices are (components are in units of the elementary charge $e$):

\begin{equation}
\begin{aligned}
Z^{(1)}_{ij} &= Z^{(4)}_{ij} = 
\begin{pmatrix}
 0.00 & 0.00 & -1.49 \\
 0.00 & 0.00 &  0.00 \\
-0.03 & 0.00 &  0.00
\end{pmatrix},\\
Z^{(2)}_{ij} &=  Z^{(3)}_{ij} = 
\begin{pmatrix}
 0.00 & 0.00 &  1.49 \\
 0.00 & 0.00 &  0.00 \\
 0.03 & 0.00 &  0.00
\end{pmatrix}.
\end{aligned}
\end{equation}

\section{Calculation of lattice sums}
We calculate lattice sum in equation (2) of the main text as follows:
\begin{multline}\label{eqSsum}
   S_{ss'}= \sideset{}{'}\sum_n\frac{e^{-i\bm{q}(\bm{R}_n+\bm{\tau}_{ss'})}}{\left|\bm{r}-(\bm{R}_n+\bm{\tau}_{ss'})\right|} = \\
    =\sum_n\frac{2}{\sqrt{\pi}}\int_0^{+\infty}e^{-i\bm{q}(\bm{R}_n+\bm{\tau}_{ss'})}e^{-\left|\bm{r}-(\bm{R}_n+\bm{\tau}_{ss'})\right|^2\eta^2}d\eta.
\end{multline} 
For $(s,s') = (1,3), (1,4), (2,3), (2,4)$, $\tau_{z,ss'}\neq0$ and making Fourier transform in eq. \eqref{eqSsum} we obtain
\begin{multline}
        S_{ss'}=\\
        \sum_{\bm{G}_m}\frac{2\pi}{\Omega_0\left|\bm{G}_m-\bm{q}\right|}e^{i(\bm{G}_m-\bm{q})\bm{r}-i\bm{G}_m\bm{\tau}_{ss'}-\left|\bm{G}_m-\bm{q}\right||z-\tau_{z,ss'}|},  
\end{multline}
which rapidly converges due to $e^{-G_m|\tau_{z,ss'}|}$.  

For $(s,s')=(1,2), (3,4)$ $\bm{\tau}_{ss'}=(\bm{\tau}_{||,ss'},0)$, so that we divide integration range by $\eta_0$, which leads to the following:
\begin{multline}\label{sum1'}
    S_{ss'}=\sum_n\frac{2}{\sqrt{\pi}}\int_0^{+\infty}e^{-i\bm{q}(\bm{R}_n+\bm{\tau}_{ss'})}e^{-\left|\bm{r}-(\bm{R}_n+\bm{\tau}_{ss'})\right|^2\eta^2}d\eta= \\
    \frac{2}{\sqrt{\pi}}\int_0^{\eta_0}\sum_n  e^{-i\bm{q}(\bm{R}_n+\bm{\tau}_{||,ss'})}e^{-\left|\bm{r}-\bm{R}_n-\bm{\tau}_{||,ss'}\right|^2\eta^2}d\eta + \\ +\frac{2}{\sqrt{\pi}}\int_{\eta_0}^{\infty}\sum_n e^{-i\bm{q}(\bm{R}_n+\bm{\tau}_{||,ss'})}e^{-\left|\bm{r}-\bm{R}_n-\bm{\tau}_{||,ss'}\right|^2\eta^2}d\eta =\\
 =\frac{2\pi}{\Omega_0}\sum_{\bm{G}_m}e^{-i\bm{G}_m\bm{\tau}_{||,ss'}+i(\bm{G}_m-\bm{q})\bm{r}}\frac{{\rm erfc}\left(\frac{|\bm{G}_m-\bm{q}|}{2\eta_0}\right)}{|\bm{G}_m-\bm{q}|}+\\
 +\sum_{n}\frac{e^{-i\bm{q}(\bm{R}_n+\bm{\tau}_{||,ss'})}{\rm erfc}(\eta_0\left|\bm{r}-\bm{R}_n-\bm{\tau}_{||,ss'}\right|)}{\left|\bm{r}-\bm{R}_n-\bm{\tau}_{||,ss'}\right|},
\end{multline}
Finally for the case $s=s'$ we obtain:
\begin{multline}\label{sum2}
  S_{ss}=\sum_n\frac{e^{-i\bm{q}\bm{R}_n}}{\left|\bm{r}-\bm{R}_n\right|}-\frac{1}{\left|\bm{r}\right|}=\\
  =\sum_n\frac{2}{\sqrt{\pi}}\int_0^{+\infty}e^{-i\bm{q}\bm{R}_n}e^{-\left|\bm{r}-\bm{R}_n\right|^2\eta^2}d\eta-\frac{1}{\left|\bm{r}\right|}= \\
=  \frac{2}{\sqrt{\pi}}\int_0^{\eta_0}\sum_n  e^{-i\bm{q}\bm{R}_n}e^{-\left|\bm{r}-\bm{R}_n\right|^2\eta^2}d\eta + \\
+\frac{2}{\sqrt{\pi}}\int_{\eta_0}^{\infty}\sum_n  e^{-i\bm{q}\bm{R}_n}e^{-\left|\bm{r}-\bm{R}_n\right|^2\eta^2}d\eta -\frac{1}{\left|\bm{r}\right|}= \\
=\frac{2\sqrt{\pi}}{\Omega_0}\sum_{\bm{G}_m}e^{i(\bm{G}_m-\bm{q})\bm{r}}\int_0^{\eta_0}\frac{1}{\eta^2}e^{-\eta^2 z^2-\frac{\left|\bm{G}_m-\bm{q}\right|^2}{4\eta^2}}d\eta+\\
+\sum_{n\neq0}\frac{e^{-i\bm{q}\bm{R}_n}{\rm erfc}(\eta_0\left|\bm{r}-\bm{R}_n\right|)}{\left|\bm{r}-\bm{R}_n\right|}-\frac{2\eta_0}{\sqrt{\pi}}.
\end{multline}
Both sums in eqs. \eqref{sum1'} and \eqref{sum2} are rapidly converged with a proper choice of $\eta_0$.  

\section{Comparison of phonon spectra in DFT and the nearest neighbour approximation}

Lattice dynamics calculations were performed with the Phonopy 
package [\hyperlink{ref:8}{8}, \hyperlink{ref:9}{9}]. The force constants were obtained from a $3 \times 3 \times 1$ supercell of the bismuth monolayer using GGA-PBE approximation with the plane wave cutoff 600 eV, $8\times8\times1$ Monkhorst-Pack k-point mesh [\hyperlink{ref:10}{10}] and energy convergence criterion 10$^{-9}$.

\begin{figure*}[t]
    \centering
    \includegraphics[width=1.0\linewidth]{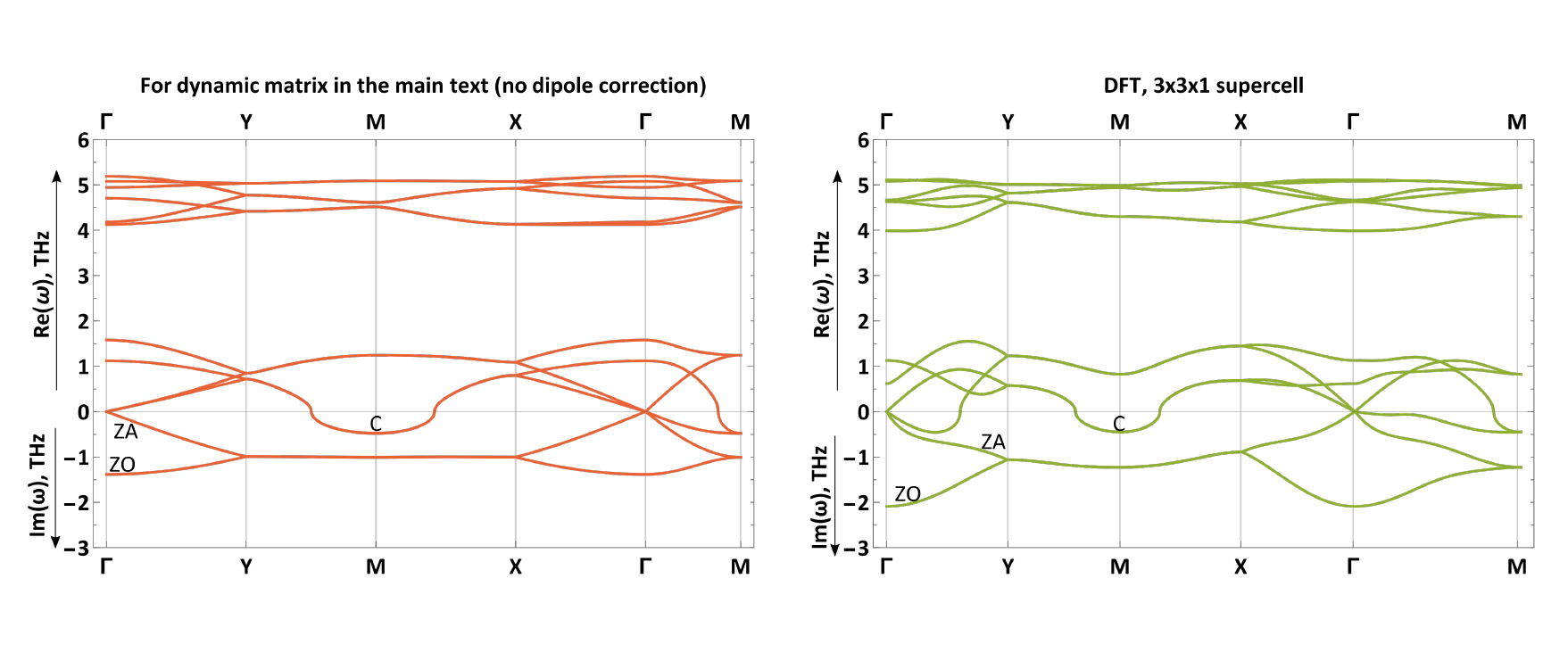}
    \caption{Phonon spectra in orthorhombic bismuth monalayers obtained with dynamical matrix from the main text (left) (without account of dipole correction) and  using the Phonopy package [\protect\hyperlink{ref:8}{8}, \protect\hyperlink{ref:9}{9}] with 3x3x1 supercell (right).}
    \label{fig:phonons}
\end{figure*}

In Fig. \ref{fig:phonons} we display all phonon branch dispersions obtained with the help of dynamical matrix in the nearest neighbour approximation for force constants from Table S1 in the main text (left panel) and that of Phonopy package (right panel) [\hyperlink{ref:8}{8}, \hyperlink{ref:9}{9}], which are in a good quantitative agreement. In both cases we used the experimental lattice parameters.

\section{Stability of phonon spectra for non-polar orthorhombic bismuth  lattice}

\begin{figure*}[t]
    \centering
    \includegraphics[width=1.0\linewidth]{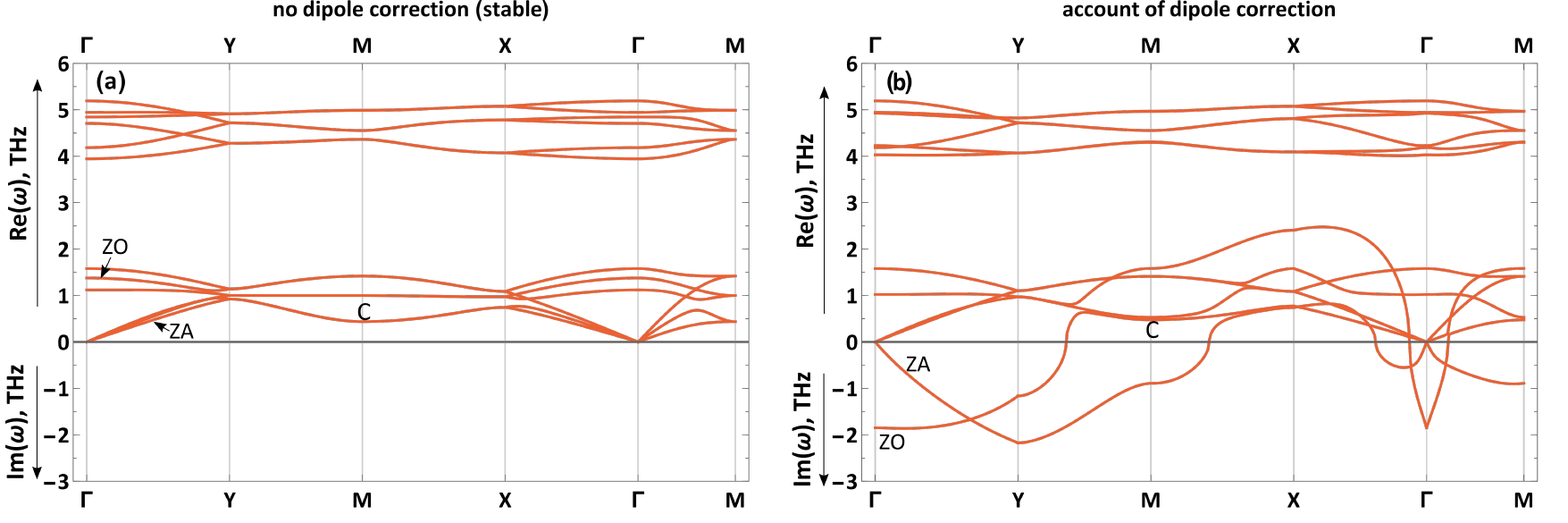}
    \caption{(a) and (b) are phonon spectra in nonpolar (symmetric lattice) orthorhombic bismuth monolayers obtained for parameters listed in Table S1 without  and with dipole-dipole interaction, respectively.}
    \label{fig:stabile_phonons}
\end{figure*}

We find (see Fig. \ref{fig:stabile_phonons}) that in the nearest neighbor approximation model with experimental lattice parameters to obtain stability for ZO optical branch (which also produces stabile dispersion for ZA branch) it is enough to change sign of $\Phi^{12}_{zz}$, compared to the initial DFT-obtained values, which also inflicts small change of $\Phi^{14}_{zz}$, due to indentity $\Phi^{14}_{zz}+\Phi^{11}_{zz}+2\Phi^{12}_{zz}=0$. In addition, 5\% decrease of $\Phi^{12}_{xy}$ makes real frequency for the chiral (C) mode in M point. Nevertheless, account of dipole-dipole interaction results in imaginary frequency for optical ZO mode, see Fig. \ref{fig:stabile_phonons}(b).

\begin{table}[htbp]
\centering
\caption{Table S1. Corrected values of parameters (highlighted by bold) of dynamical matrix (eV/\AA$^2$) for stability of phonon spectra without dipole-dipole interaction.  \label{tab}}
\begin{tabular}{|l|c|c|c|c|}
\hline
 & \multicolumn{1}{c|}{\(xx\)} & \multicolumn{1}{c|}{\(yy\)} & \multicolumn{1}{c|}{\(zz\)} & \multicolumn{1}{c|}{\(xz\)/\(xy\)/\(yz\)} \\
 \hline
\(\Phi^{11}\) & 8.348 & 11.505 & 9.1309 & 1.689 (\(xz\)) \\
\(\Phi^{12}\) & -3.74 & -5.22 & {\bf -0.409} & {\bf 4.313} (\(xy\)), -0.13 (\(yz\)) \\
\(\Phi^{14}\) & -0.868 &  -1.065 & {\bf -8.321} & --- \\
\hline
\end{tabular}
\end{table}

\section{Phonon spectra in ferroelectric phase of orthorhombic bismuth monolayers}

\begin{figure*}[t]
    \centering
    \includegraphics[width=1.0\linewidth]{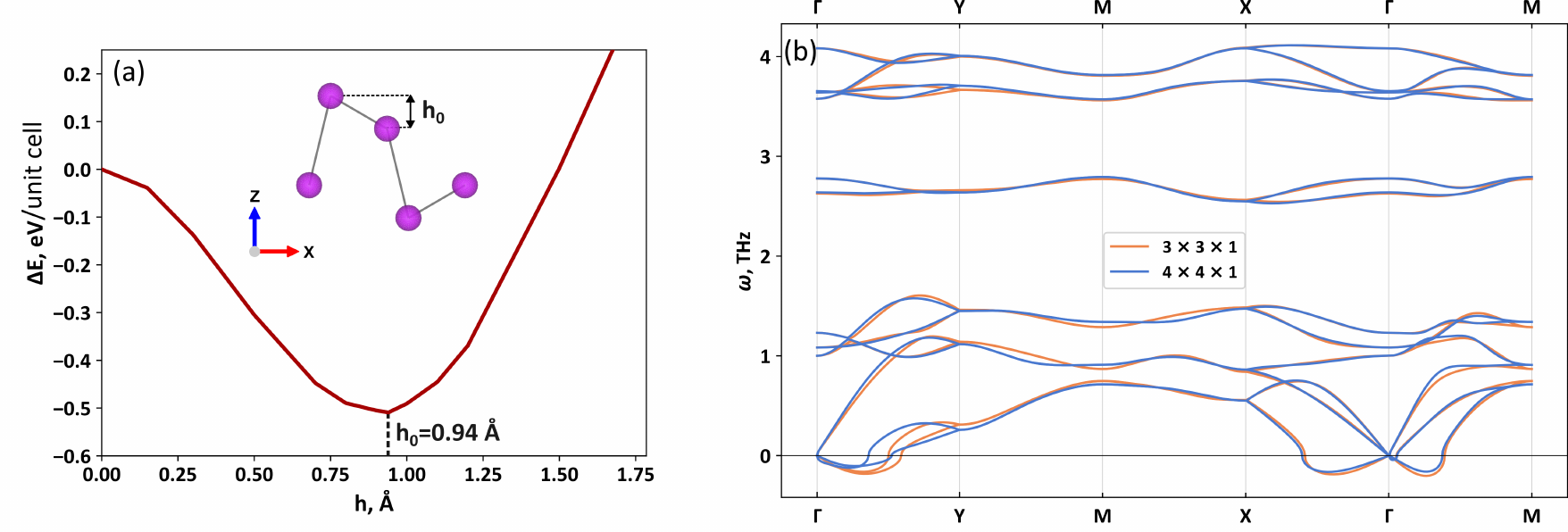}
    \caption{(a) Energy difference per unit cell between ferroelectric and non-polar structures of orthorhombic bismuth monolayers. Inset shows unit cell in the ferroelectric phase.  (b) Phonon spectra for ferroelectric orthorhombic bismuth monolayers (shown on inset on panel (a)), obtained for $3\times3\times1$ and $4\times4\times1$ supercells.}
    \label{fig:phonons_in_Ferro}
\end{figure*}

In Fig. \ref{fig:phonons_in_Ferro}(a) we find optimal value of ZO mode amplitude, $h_0=0.94$\,\AA, that provides gain in energy for the ferroelectric phase orthorhombic bismuth monolayers with respect to the non-polar lattice with experimental lattice parameters $a=4.75$\,\AA, $b=4.42$\,\AA, and $d=2.8$\,\AA. Calculated in DFT phonon spectra (Fig. \ref{fig:phonons_in_Ferro}(b)) with $3\times3\times1$ and $4\times4\times1$ supercells show that lattice structure of the ferroelectric phase is stable, as all optical phonons possess positive energies, whereas small imaginary values for acoustic phonons near $\Gamma$ point are attributed to numerical accuracy related to supercell sizes [\hyperlink{ref:11}{11}].

\vspace{0.5em}
\hrule
\vspace{0.5em}

\begin{list}{[\arabic{enumi}]}{\usecounter{enumi}\setlength{\leftmargin}{2.5em}\setlength{\labelwidth}{2em}\setlength{\labelsep}{0.5em}\setlength{\itemsep}{0pt}\setlength{\parsep}{0pt}}
\item \hypertarget{ref:1}{G. Kresse, J. Hafner, Phys. Rev. B \textbf{47}, 558 (1993).}
\item \hypertarget{ref:2}{G. Kresse and J. Furthmüller, Phys. Rev. B \textbf{54}, 11169 (1996).}
\item \hypertarget{ref:3}{D. M. Ceperley and B. J. Alder, Phys. Rev. Lett. \textbf{45}, 566 (1980).}
\item \hypertarget{ref:4}{J. P. Perdew, K. Burke, and M. Ernzerhof, Phys. Rev. Lett. \textbf{77}, 3865 (1996).}
\item \hypertarget{ref:5}{P. E. Blöchl, Phys. Rev. B \textbf{50}, 17953 (1994).}
\item \hypertarget{ref:6}{J. Neugebauer and M. Scheffler, Phys. Rev. B \textbf{46}, 16067 (1992).}
\item \hypertarget{ref:7}{J. Gou, L. Kong, X. He, Y. L. Huang, J. Sun, S. Meng, K. Wu, L. Chen, A. T. S. Wee, Sci. Adv. \textbf{6}, eaba2773 (2020).}
\item \hypertarget{ref:8}{A. Togo, I. Tanaka, Scr. Mater. \textbf{108}, 1 (2015).}
\item \hypertarget{ref:9}{A. Togo, J. Phys. Soc. Jpn. \textbf{92}, 012001 (2023).}
\item \hypertarget{ref:10}{H. J. Monkhorst, J. D. Pack, Phys. Rev. B \textbf{13}, 5188 (1976).}
\item \hypertarget{ref:11}{E. Aktürk, O. Ü. Aktürk, S. Ciraci, Phys. Rev. B \textbf{94}, 014115 (2016).}
\end{list}

\end{document}